\documentclass[a4paper,aps,superscriptaddress,twocolumn,nofootinbib]{revtex4-1}

\usepackage{graphicx}
\usepackage{amsmath,amsthm,amssymb,bbm}
\newcommand{\bra}[1]{\langle#1|}
\newcommand{\ket}[1]{|#1\rangle}
\newcommand{\proj}[1]{|#1\rangle\!\langle#1|}

\newcommand{\Tr}{\operatorname{Tr}}

\newcommand{\identity}{\mathbbm{1}}

\begin{document}

\title{Optimal randomness certification from one entangled bit}

\author{Antonio~Ac\'in}
\affiliation{ICFO-Institut de Ciencies Fotoniques, Mediterranean
Technology Park, 08860 Castelldefels (Barcelona), Spain}
\affiliation{ICREA-Instituci\'o Catalana de Recerca i Estudis Avan\c cats, Lluis Companys 23, 08010 Barcelona, Spain\\
}
\author{Stefano Pironio}
\affiliation{Laboratoire d'Information Quantique, Universit\'e libre de Bruxelles (ULB), 1050 Brussels, Belgium}

\author{Tam\'as V\'ertesi}
\affiliation{Institute for Nuclear Research, Hungarian Academy of Sciences, H-4001 Debrecen, P.O. Box 51, Hungary}

\author{Peter Wittek}
\affiliation{ICFO-Institut de Ciencies Fotoniques, Mediterranean
Technology Park, 08860 Castelldefels (Barcelona), Spain}
\affiliation{University of Bor{\aa}s, 50190 Bor{\aa}s, Sweden}

\date{May 14, 2015}

\begin{abstract}
By performing local projective measurements on a two-qubit entangled state one can certify in a device-independent way up to one bit of randomness. We show here that general measurements, defined by positive-operator-valued measures, can certify up to two bits of randomness, which is the optimal amount of randomness that can be certified from an entangled bit. General measurements thus provide an advantage over projective ones for device-independent randomness certification.
\end{abstract}

\maketitle

The non-local correlations observed when measuring entangled quantum particles certify the presence of intrinsic randomness in the measurement outputs in a way that is independent on the underlying physical realization of these correlations.  While this relation between nonlocality and randomness had been noted by different authors since the seminal work by Bell~\cite{bell,rmpbell}, it is only recently that the tools to quantify the intrinsic randomness produced in Bell setups were provided~\cite{nature,njp,scaranibancal}. These tools were initially introduced in the context of device-independent randomness generation~\cite{colbeck,nature,vv,millershi}, but have also allowed us to obtain a much better understanding of the relation between randomness and Bell violations, two of the most fundamental properties of quantum theory. For instance, today we know that maximal randomness can be certified from arbitrarily small amounts of non-locality or entanglement~\cite{amp}, or that maximal randomness certification is possible in quantum theory, but not in general theories restricted only by the no-signalling principle~\cite{maxrand}.

Despite all this progress, there are still fundamental questions on the relation between randomness, non-locality, and entanglement that remain completely unexplored. In this work we consider and solve one of them: we obtain the maximal amount of randomness that can be certified in a standard Bell scenario involving local measurements on one entangled bit or $\emph{ebit}$. In order to achieve this maximum the use of general measurement beyond projective ones, often known as Positive-Operator-Valued Measures (POVM), is necessary. Thus, our results and techniques are also interesting because they provide one of the few examples in the context of Bell non-locality where the use of these general measurements provides an advantage over standard projective measurements (other examples can be found in~\cite{gisin,tamas}).

We formulate the relation between randomness and non-locality in the setting of \emph{non-local guessing games} as considered in \cite{njp}. Such games consist of two users, Alice and Bob, and an adversary Eve. Alice and Bob perform local measurements on two separate quantum systems, labelled by $A$ and $B$. There are $m_A$ and $m_B$ possible measurements on particles $A$ and $B$, each producing $r_A$  and $r_B$ possible results. Measurement choices are labelled by $x$ and $y$, with $x=1,\ldots,m_A$ and $y=1,\ldots,m_B$, while the corresponding results are labelled by $a$ and $b$, with
$a=1,\ldots,r_A$ and $b=1,\ldots,r_B$, respectively. The behavior of Alice and Bob's systems is characterized by the finite set of $m_A\times m_B\times r_A\times r_B$ probabilities $P=\{P(ab|xy)\}$, where $P(ab|xy)$ is the probability that outcomes $a$ and $b$ are obtained when performing measurements $x$ and $y$ on particles $A$ and $B$. In our non-local guessing game, $P$ is assumed to be given, i.e. it is a promise on the behavior of Alice's and Bob's systems. The aim is for Eve to guess as well as possible Alice's outcome for a certain input $\bar x$. To achieve this, Eve can prepare Alice's and Bob's system in any way compatible with the given behavior $P$ and the laws of quantum physics. A strategy $S$ for Eve consists in $i)$ a tripartite quantum state $|\Psi\rangle_{ABE}$ on a composite Hilbert space  $\mathcal{H}_A\otimes \mathcal{H}_B\otimes\mathcal{H}_E$ of arbitrary dimension characterizing the possible correlations between Alice's, Bob's and Eve's system, $ii)$ for each value of $x$, a POVM $A_x$ on $\mathcal{H}_A$ with elements $A_{a|x}$ and for each value of $y$, a POVM $B_y$ on $\mathcal{H}_B$ with elements $B_{b|y}$ characterizing the local measurements of Alice and Bob, $iii)$ a POVM $Z$ on $\mathcal{H}_E$ with elements $Z_a$ whose result is Eve's best guess on Alice's outcome. Such a strategy is compatible with $P$ if
\begin{equation}
\label{qcorr}
P(ab|xy)=\bra{\Psi_{ABE}}A_{a|x}\otimes B_{b|y}\otimes I\ket{\Psi_{ABE}}.
\end{equation}
The figure of merit of the game is the probability
that Alice's output and Eve's guess coincide, maximized over all strategies $S$ compatible with $P$ (a set which we denote $\mathcal{S}_P$):
\begin{equation}
\label{locguess}
G(\bar x,P)=\max_{S\in \mathcal{S}_P}\sum_a \bra{\Psi_{ABE}}A_{a|\bar x}\otimes I	 \otimes Z_a\ket{\Psi_{ABE}}.
\end{equation}
We refer to this quantity as the \emph{local guessing probability}.

We can also introduce a variant of the game in which Eve attempts to guess both Alice's and Bob's outputs for a given pair of inputs $(\bar x, \bar y)$ in which case her strategies involve a POVM $Z$ with elements $Z_{ab}$ and the figure of merit is
\begin{equation}
\label{gloguess}
G(\bar x,\bar y,P)=\max_{S\in \mathcal{S}_P}\sum_{ab} \bra{\Psi_{ABE}} A_{a|\bar x}\otimes B_{b|\bar y} \otimes Z_{ab}\ket{\Psi_{ABE}}.
\end{equation}
We refer to this quantity as the \emph{global guessing probability}.

The local and global guessing probabilities quantify the predictability of the result of measurement $\bar x$, or of pair of measurements $\bar x$ and $\bar y$, by a quantum observer with an optimal description of the experiment. Taking minus the logarithm in base two of these quantities gives a measure of randomness expressed in bits. A bound on these quantities is often a central element in the analysis of actual randomness generation of expansion protocols, such as \cite{nature,vv}, where it directly determines (up to statistical corrections) the final amount of randomness generated.

Note that we always have $G(\bar x,P)\geq \max_a P(a|\bar x)$ and $G(\bar x,\bar y,P)\geq \max_{ab} P(ab|\bar x\bar y)$ since a simple strategy is for Eve to simply guess the most probable outcomes of Alice's and Bob's measurements without exploiting any further detailed information about Alice's and Bob's systems. However, in general a non-trivial strategy performs strictly better than these trivial bounds. Note that the guessing probabilities satisfy a convexity property in the sense that if $P$ admits the convex decomposition $P=\sum_\lambda q_\lambda P_\lambda$ in term of quantum realizable behaviors $P_\lambda$, then $G(\bar x,P)\geq \sum_\lambda q_\lambda G(\bar x,P_\lambda)$ and similarly for $G(\bar x,\bar y, P)$. This follows from the fact that a strategy that Eve can follow is to prepare Alice's and Bob's system with probability $q_\lambda$ according to the behavior $P_\lambda$ and use the optimal guessing strategy associated to $P_\lambda$.
In particular in the case in which the given correlations $P$ can be described by a local model, i.e. can be written down as a convex mixture of deterministic behaviors, one has $G(x)=1$ and $G(x,y)=1$ for any measurement $x$ and $y$. However, the violation of Bell inequalities does not necessarily imply that $G(\bar x)<1$ or $G(\bar x,\bar y)<1$.
Upper-bounds on the guessing probabilities can be computed using the Navascues-Pironio-Acin (NPA) hierarchy for quantum correlations~\cite{npa}, as shown in \cite{njp,scaranibancal}.

In this work, our goal is to compute the maximal randomness that one can certify from one ebit. That is, our goal is to identify the correlations minimising the guessing probability among all those attainable by measuring an entangled state equal in its Schmidt form to $|\phi^+\rangle=(|00\rangle+|11\rangle)/\sqrt{2}$. The obtained quantity defines the optimal amount of randomness that can be certified in a device-independent way using an entangled qubit.

Local measurements on such a state can always be viewed as POVMs acting on a qubit since the local Schmidt dimension is 2. In the case where the local qubit measurements are projective, it is known that 1 bit of local randomness \cite{nature} and 2 bits of global randomness \cite{acin,scarani} can be certified from an ebit. This is also the maximum that can be achieved under such measurements, since a qubit projective measurement has only two possible outcomes. Beating those bounds thus requires considering more general measurements, beyond projective.

We start by stating a rather straightforward observation: no more than $2\log d$ bits of local randomness and $4\log d$ bits of global randomness can be certified by measuring an entangled state of dimension $d\times d$. This follows from the convexity property of the guessing probabilities mentioned above and the fact that a POVM acting on a space of dimension $d$ can always be decomposed as a convex sum of POVMs of at most $d^2$ outputs~\cite{extrpovm}, which can evidently contain at most $2\log d$ bits of randomness. In the case of qubits, no more than 2 bits of local randomness and 4 bits of global randomness can be certified, i.e. twice as much than using projective measurements.

Our main result is to construct two examples of qubit correlations saturating this bound on the local randomness. They thus provide examples of optimal randomness certification from one ebit. In the first example we prove analytically that the local randomness is 2 bits, while in the second we have to resort to semidefinite programming (SDP) techniques.

\textit{First optimal construction}--- Our first construction is based on non-local correlations obtained by measuring the two-qubit maximally entangled state $|\phi^+\rangle$ with measurements $x=1,2,3$ on Alice's side corresponding to $\sigma_x,\sigma_y,\sigma_z$ and with measurements $y=1,\ldots,6$ on Bob's side corresponding to $(\sigma_x\pm\sigma_y)/\sqrt{2}, (\sigma_x\pm\sigma_z)/\sqrt{2},(\sigma_y\pm\sigma_z)/\sqrt{2}$. These measurements are chosen so that they produce the maximal violation of the Clauser-Horne-Shimony-Holt (CHSH) inequality~\cite{chsh} with all the possible pairs of measurements on the first particle, that is $\mathcal{B}(1,2;1,2	)=\mathcal{B}(1,3;3,4)=\mathcal{B}(2,3;5,6)=2\sqrt{2}$, where
\begin{equation}
\label{CHSH}
\mathcal{B}(i,j;k,l)=E_{ik}+E_{il}+E_{jk}-E_{jl}\nonumber
\end{equation}
and $E_{ik}=\sum_{ab}(-1)^{ab}P(ab|ik)$.
Finally a four-output measurement $y=7$ is included on the second particle. This seventh measurement by Bob plays a special role in our construction as it is the one used to certify the two random bits. We denote it by $R$, for random, and its measurement operators by $R_b$, with $b=1,\ldots,4$. This measurement is rather generic but has to satisfy two requirements: (i) it is extremal in the set of qubit measurements and (ii) its measurement operators are such that $\Tr(R_b)=1/4,\,\forall b$. An example of such a measurement is given by a POVM where $R_b=1/4\proj{\psi_b}$ and the Bloch vectors corresponding to $\ket{\psi_b}$ point to the direction of a tetrahedron.

The measurements described above define the behavior $P$ which Eve has to reproduce (possibly using other quantum realizations of arbitrary dimension). Clearly,  the four-output measurement $R$, when acting on half of a maximally entangled state, gives $P(b|y=7)=1/4$ for all $b$ and thus $G(y=7,P)\geq 1/4$. As mentioned previously, this is only a trivial upper bound on the amount of intrinsic randomness, which is generally far from being tight. However, we now prove that for the correlations $P$ defined above the bound is tight and, therefore, $G(y=7,P)=1/4$ and hence two-bits of local randomness can be certified from $P$.

To understand the main intuition behind the choice of state and measurements in our construction, the idea is to exploit the fact that, roughly speaking, the only quantum way of getting the maximal quantum violation of the CHSH inequality, also known as the Tsirelson's bound, is by performing anti-commuting measurements on a two-qubit maximally entangled state. We refer to these quantum correlations, which are unique, as Tsirelson correlations. The correlations generated in the previous quantum setup contain three blocks of Tsirelson correlations for the different pairs of settings on particle $A$ and corresponding measurements on $B$. This suggests that the only state and measurements that could have produced these correlations should be, up to local unitary transformations, precisely Pauli measurements $X$, $Y$ and $Z$ on $A$ acting on a two-qubit maximally entangled state. Now, these three measurements when acting on half of a maximally entangled state allow reconstructing any measurement implemented on the other half. In fact they remotely project particle $B$ onto the eigenstates of these three observables, which are tomographically complete.  Therefore, it should be possible from the observed correlations to reconstruct and certify the POVM elements implemented on $B$ and conclude that they certify the desired amount of randomness. As we will see, this intuition is not entirely true due to a problem with complex conjugation, but it is enough to prove the desired result. In fact, we believe this construction is interesting per se and may find applications in other problems of certification and self-testing of quantum devices.

After providing this intuition, let us now show that the given setup indeed certifies two bits of local randomness. As described earlier, any quantum strategy of Eve corresponds to a tripartite state $|\Psi\rangle_{ABE}$, a set of three observables $A_1,A_2,A_3$ for Alice, six two-output observables $B_1,\ldots,B_6$ and one four-output measurement $B_7$. This strategy should reproduce the given correlations $P$, as expressed (\ref{qcorr}), and thus also the CHSH expectations $\mathcal{B}(1,2;1,2	 )=\mathcal{B}(1,3;3,4)=\mathcal{B}(2,3;5,6)=2\sqrt{2}$. This implies, given the self-testing property of the CHSH inequality \cite{stchsh} and following Mosca-McKague \cite{mosca-mckague}, that up to a local isometry $|\Psi_{ABE}\rangle = |\phi^+\rangle_{AB}|\psi\rangle_{A'B'E}$. In addition $A_1|\Psi_{ABE}\rangle = ({X}_A\otimes I_{A'}) |\phi^+\rangle_{AB}|\psi\rangle_{A'B'E}$ and similarly $A_3|\Psi_{ABE}\rangle = (Z_A\otimes I_{A'}) |\phi^+\rangle_{AB}|\psi\rangle_{A'B'E}$. On the other hand, $A_2|\Psi_{ABE}\rangle = (Y_A\otimes M_{A'}) |\phi^+\rangle_{AB}|\psi\rangle_{A'B'E}$ where $M_{A'}$ is hermitian and unitary. Basically this states that the three observables $A_1,A_2,A_3$ are necessarily the Pauli measurements $X$, $Y$, $Z$ acting on Alice's system, except for $A_2$ for which a correction $M_{A'}$ is needed. This correction reflects the fact that the optimal measurements leading to the three maximal CHSH expectations given above are only defined up to a complex conjugation (see \cite{mosca-mckague} for a	 discussion of this point).

Let us now determine the action of the POVM $B_7$. For simplicity of notation, let us denote it $\tilde R$ and the corresponding outcome operators $\tilde R_b$.
The correlations between the outcomes of this POVM and Alice's observables should equal those of the ideal set-up defined earlier, as expressed in (\ref{qcorr}). This means that $\langle \Psi_{ABE}|A_\mu\otimes \tilde R_b\otimes I|	 \Psi_{ABE}\rangle=\langle \phi^+|\sigma_\mu\otimes R_b|\phi^+\rangle$, where $A_0$ denotes the identity operator.

Let us now note that the ideal POVM elements $R_b$ used in the definition of $P$ can be written as $R_b=\sum_\mu r_b^\mu \sigma_\mu$ where $\{\sigma_\mu : \mu=0,1,2,3\}$ is the basis of the four Pauli operators and $r_b^\mu$ are complex coefficients defining the POVM. We then have $\langle\phi^+|\sigma_\mu\otimes R_b|\phi^+\rangle=r_b^\mu$, hence $\langle \Psi_{ABE}|A_\mu\otimes \tilde R_b\otimes I|	\Psi_{ABE}\rangle=r_b^\mu$.

On the other hand, since $\tilde R_B$ acts jointly on systems $BB'$, without loss of generality we can decompose its operators as $\tilde R_b = \sum_\mu \sigma_\mu \otimes \tilde R_b^\mu$, where $\{\sigma_\mu : \mu=0,1,2,3\}$ is the basis of the four Pauli operators on $B$ and $\tilde R_b^\mu$ are arbitrary hermitian operators on $B'$. Inserting these expressions for $\tilde R_b$ and using the specific form of $|\Psi_{ABE}\rangle$ and $A_1,A_2,A_3$ enforced by the CHSH constraints, we find
$\langle \psi_{A'B'E}|I\otimes \tilde R_b^\mu\otimes I|\psi_{A'B'E}\rangle=r_b^\mu$ in the case $\mu\neq 2$ and
$\langle \psi_{A'B'E}|M_{A'}\otimes \tilde R_b^\mu\otimes I|\psi_{A'B'E}\rangle=r_b^\mu$ when $\mu=2$.

Introduce the normalized states $|\varphi^{\pm,e}_{B'}\rangle=(M^{\pm}_{A'}\otimes I\otimes Z_e)|\psi_{A'B'E}\rangle/\sqrt{q_{\pm,e}}$, where $M^{\pm}_{A'}$ is the projector on the $\pm$ eigenspace of $M_{A'}$ and $Z_e$ is the projector corresponding to Eve's outcome $e$ (without generality we can assume Eve's measurement $Z$ to be projective). We can then write
\begin{eqnarray}
r^\mu_b&=&\sum_{\pm,e} q_{\pm,e}	\langle\varphi^{\pm,e}|I\otimes \tilde R^\mu_b\otimes I|\varphi^{\pm,e}\rangle\nonumber\\
&=&\sum_{e} \left[q_{+,e}\, \tilde r_{b}^{\mu;+,e}+q_{-,e}\, \tilde r_{b}^{\mu;-,e}\right]\label{povm1}
\end{eqnarray}
for $\mu=0,1,3$
and
\begin{eqnarray}
r^\mu_b&=&\sum_{\pm,e} q_{\pm,e}	\langle\varphi^{\pm,e}|M_{A'}\otimes \tilde R^\mu_b\otimes I|\varphi^{\pm,e}\rangle\nonumber\\
&=&\sum_{e} \left[q_{+,e}\, \tilde r_{b}^{\mu;+,e}-q_{-,e} \tilde r_{b}^{\mu;-,e}\right]\label{povm2}
\end{eqnarray}
for $\mu=2$, where we have defined the coefficients $\tilde r_{b}^{\mu;\pm,e}=\langle\varphi^{\pm,e}|I\otimes R^\mu_b\otimes I|\varphi^{\pm,e}\rangle$. Note that these coefficients define a family of valid qubit POVMs $\tilde R^{\pm,e}$ with operators $\tilde R_b^{\pm,e}=\sum_\mu \tilde r_{b}^{\mu;\pm,e}\sigma_\mu$. These POVMs simply correspond to preparing an ancilla system $B'$ in the state $|\varphi_{B'}^{\pm,e}\rangle$ and performing the POVM $\tilde R$ on the joint system $BB'$.

Now redefine $\tilde r_{b}^{\mu;-,e}$ as above but with the sign of $\tilde r^{2;-,e}_{b}$ changed. Then this also defines valid POVMs, which are just the the complex conjugates of $\tilde R^{\pm,e}$. With this redefinition, we can now write eqs. (\ref{povm1}) and (\ref{povm2}) as
\begin{eqnarray}
r^\mu_b&=&\sum_{\pm,e} q_{\pm,e}\, \tilde r_{b}^{\mu;\pm,e}
\end{eqnarray}
for $\mu=0,1,2,3$. We can interpret this as providing a convex decomposition for the ideal POVM $R$ in term of the POVMs $\tilde R^{\pm,e}$ with respective weights $q_{\pm,e}$. But since this POVM is extremal, we must have $\tilde r_{b}^{\mu;\pm,e}=r^\mu_b$ for all $\pm,e$. In particular $\tilde r_{b}^{0;\pm,e}=r^0_b=1/4$ for all $\pm,e$.

Finally, let us now rewrite the guessing probability for the input $y=7$ with these notations. We have
\begin{eqnarray}
G(y=7,P)&=&\sum_{b=0}^3 \langle \Psi_{ABE}|I\otimes \tilde R_b\otimes Z_b|\Psi_{ABE}\rangle \nonumber\\
&=&\sum_{b=0}^3 \langle \psi_{A'B'E}|I\otimes \tilde R^0_b\otimes Z_b|\psi_{A'B'E}\rangle \nonumber\\
&=& \sum_{b=0}^3 q_{\pm,b}\, \tilde r_{b}^{0;\pm,b}\nonumber =1/4,
\end{eqnarray}
which provide the two announced random bits.

\textit{Second optimal construction}--- Our second construction to generate two random bits from a qubit is slightly simpler but the certification of randomness makes use of the numerical SDP techniques introduced in~\cite{njp} based on the NPA hierarchy for quantum correlations~\cite{npa}.

The construction is based on the elegant Bell inequality introduced in~\cite{gisin}. It is defined in a scenario involving three measurements on Alice's side and four on Bob's. All measurements have two outputs and the inequality reads
\begin{eqnarray}
\label{elineq}
\beta_\text{el}&=&E_{11}+E_{12}-E_{13}-E_{14}+E_{21}-E_{22}+E_{23}-E_{24}\nonumber\\
&+&E_{31}-E_{32}-E_{33}+E_{34}\leq 6 .
\end{eqnarray}
The maximal known quantum violation of the inequality is equal to $4\sqrt 3$ and is obtained with a maximally entangled state, projective measurements $A_1=\sigma_x$, $A_2=\sigma_y$, $A_3=\sigma_z$ on Alice, while Bob's projective measurements are defined by the four vectors of a tetrahedron. In Appendix A, we show that this known quantum violation is in fact optimal, which gives a new Tsirelson-type bound for quantum correlations.

We introduce a four outcome measurement $R$, but now on Alice's side. As above, this measurement will be used to generate the two random bits. Given these configuration of measurements, we define the modified elegant Bell inequality
\begin{equation}
\label{model}
\beta'_\text{el}=\beta_\text{el}-k\sum_{i=1}^4P(a=i,b=+1|x=4,y=i)\leq 6 ,
\end{equation}
where $k$ is an arbitrary strictly positive constant. As the last term in the inequality is always negative, the bound follows from the bound on $\beta_\text{el}$. The same argument implies that the quantum violation cannot be larger than $4\sqrt 3$. If we use the known optimal qubit settings, given above, the only way of getting this maximal violation is if the POVM elements of measurement $R$ are anti-aligned with the four projective measurements on Bob's side, so that all probabilities $P(a=i,b=+1|x=4,y=i)$ are zero. But then, the corresponding measurement, when acting on half of a maximally entangled state, define two random bits on Alice's side. The intuition, then, is that the maximal violation of the modified elegant Bell inequality should certify the generation of two random bits for measurement $R$.

We used the numerical techniques in~\cite{njp} to bound the randomness present in the correlations maximally violating~\eqref{model}. Recall that these techniques are based on SDP and, therefore, one has control over the precision of the numerical result. Using these techniques at order $2+AAB+ABB$ with an arbitrary-precision solver~\citep{ncpol2sdpa,sdpagmp}, we can show that the generated randomness by measurement $R$ in the previous setup is larger than $1.99999989474702$ bits.

\begin{figure}[tb!]
\centering
\includegraphics[width=\columnwidth]{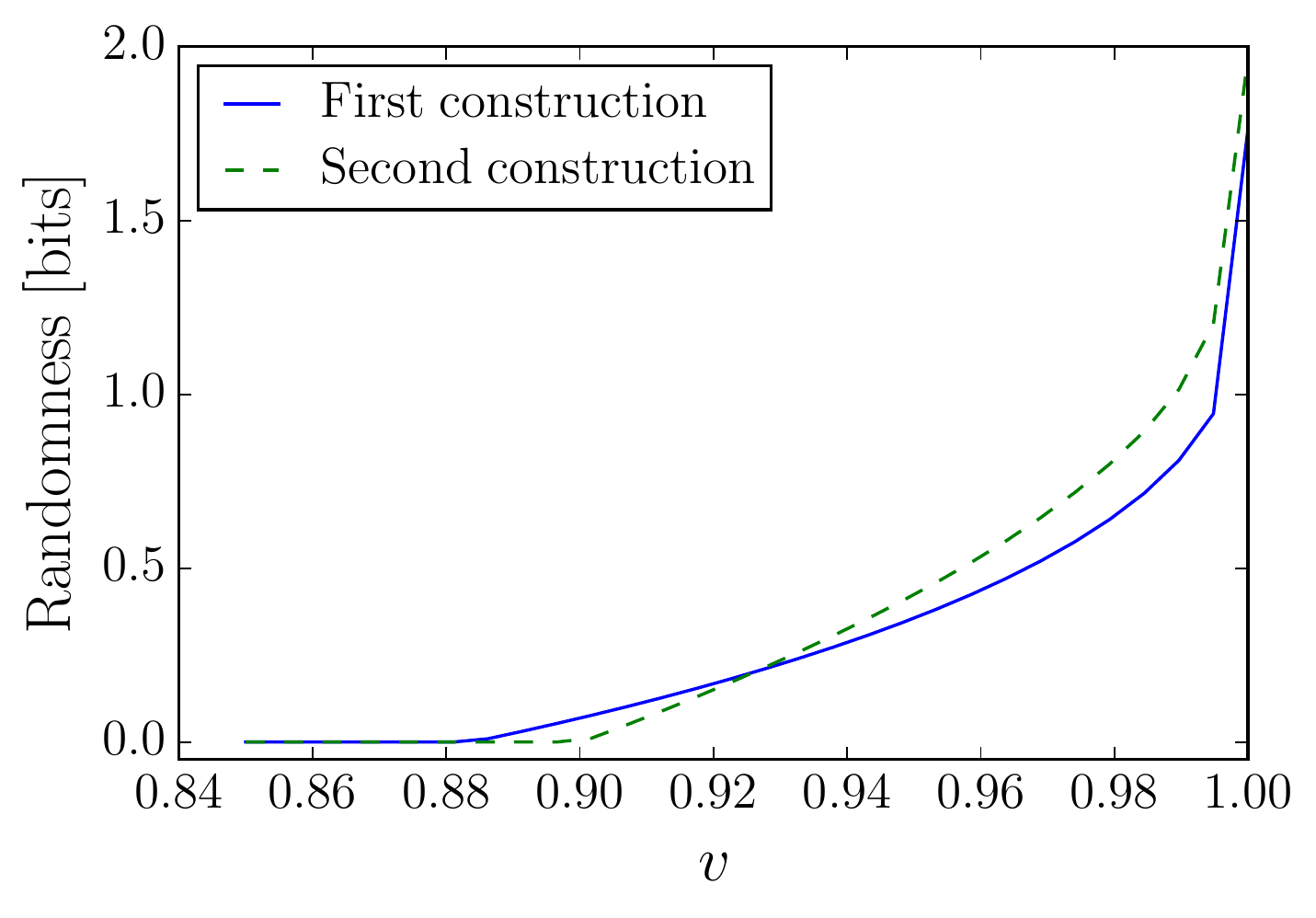}
\caption{Lower bound on the randomness ($-\log_2$ of the guessing probability) as a function of the visibility $v$ ranging from 0.85 to 1.0. The correlations are of the form $v\mathbf{q} + (1 - v)\mathbf{r}$, where $\mathbf{q}$ are the quantum correlations yielding the maximum violation of the respective inequalities, and $\mathbf{r}$ denotes the completely unbiased correlations that assign the same probability to each measurement outcome. The figure was obtained using level 2 of the NPA hierarchy and thus only represents a lower bound on the maximal randomness (note for instance that we do not recover the optimal values of 2 bits for $v=1$). Better noise resistance than the one provided by these curves may thus be obtained by performing a more complex analysis.   
}
\label{noise_sensitivity}
\end{figure}

\textit{Noise robustness}--- The numerical approach of~\cite{njp} also allows one to study the robustness of the previous constructions against noise. A typical noise model consists of mixing, with weights $v$ and $1-v$, with $0\leq v\leq 1$, the ideal quantum correlations with uncorrelated noise in which all outputs have the same probability. By decreasing $v$, often known as visibility, the amount of certified randomness decreases. In Figure~\ref{noise_sensitivity} we plot a lower bound on the generated randomness as a function of the visibility for the two previous constructions. It can be seen that the gain provided by the POVM is fragile, in the sense that a small fraction of noise, of the order of 0.01, makes the obtained randomness smaller than one bit, which is the randomness provided by projective measurements. These considerations are relevant, for instance, when thinking of a possible experiment showing the advantage of using POVM's for randomness certification. A natural open question opened by our work is thus to identify robust setups for randomness generation using POVM's.

\textit{Global randomness}--- Before concluding, we would like to briefly discuss the problem of global randomness. The question is whether it is possible to find Bell setups involving a pair of maximally entangled qubits allowing the generation of $4$ bits of randomness. We mainly leave this question for future work. Nevertheless, we made some preliminary numerical searches, using slightly more complex variations of the previous construction based on the elegant Bell inequality. These constructions are described in Appendix B. As shown there, they can be used to certify the generation of 2.8997 bits, which is both higher than the global randomness that can be certified with projective measurements and the local randomness that can be certified with general measurements.

\textit{Conclusions}---
We have shown that an ebit can certify the presence of more than one bit of randomness locally and more than two bits globally. For the case of local randomness, we have found two constructions that can certify 2 bits of randomness, the maximal possible value. Both constructions involve a maximally entangled state and the three Pauli measurements on Alice's side. In the first construction, the violation of three CHSH inequalities are used to self-test the maximally entangled state and these three Pauli measurements on Alice's side. This in turn allows to self-test a four-outcome extremal POVM on Bob's, which generates the two bits of local randomness. Our second construction is instead based on a single inequality -- the elegant inequality introduced in \cite{gisin}. In this case, we had to resort to SDP techniques to put a bound on the local randomness. A possible way to proof this bound analytically, would be to first proof that the elegant Bell inequality also provides a self-test for the maximally entangled state and the three Pauli measurements on Alice's side. It would then be possible to add the four-outcome measurement on Bob's side and use the same proof as above to conclude that it generates two random bits.

Both constructions are quite sensitive to noise, as a fraction of noise of the order of 0.01 makes the obtained randomness smaller than one bit, which can already be obtained with more robust constructions based on projective measurements. A natural open question is thus to identify robust and optimal setups for randomness certification using POVM's.

Finally, we also found a construction based on POVMs which yields more than two bits of global randomness, the best that can be obtained with projective measurements, but less than the theoretical maximum of four bits. It remains an open question whether this maximum can actually be attained.

\textit{Acknowledgements}---We acknowledge financial support from the EU projects QALGO and SIQS, the ERC CoG QITBOX, the F.R.S.-FNRS under the project DIQIP, the Brussels-Capital Region through a BB2B grant, the Spanish project FOQUS, the Generalitat de Catalunya (SGR875), the Hungarian National Research Fund OTKA (K111734), the J\'anos
Bolyai Programme of the Hungarian Academy of Sciences and the John Templeton Foundation. S. P. is a Research Associate of the Fonds de la Recherche Scientifique F.R.S.-FNRS (Belgium).

\appendix

\section{SOS decomposition of the elegant Bell inequality}
The aim of this Appendix is to prove a tight upper bound on the quantum violation of the elegant Bell inequality $\beta_\text{el}$ defined by (\ref{elineq}) in the main text.

Using the maximally entangled two-qubit state $|\phi^+\rangle=(|00\rangle+|11\rangle)/\sqrt{2}$ and projective measurements $A_1=\sigma_x$, $A_2=\sigma_y$, $A_3=\sigma_z$ on Alice's side, while the following four projective measurements on Bob's side
\begin{align}
B_1 &= (\sigma_x-\sigma_y+\sigma_z)/\sqrt 3 \nonumber\\
B_2 &= (\sigma_x+\sigma_y-\sigma_z)/\sqrt 3 \nonumber\\
B_3 &= (-\sigma_x-\sigma_y-\sigma_z)/\sqrt 3 \nonumber\\
B_4 &= (-\sigma_x+\sigma_y+\sigma_z)/\sqrt 3,
\label{tetra}
\end{align}
we obtain the Bell violation $4\sqrt 3$.

We now show that this amount of Bell violation for inequality $\beta_\text{el}$ is optimal. To this end, let us redefine the Bell operator, $\bar{\beta}_\text{el}\equiv 4\sqrt 3\identity - \hat\beta_\text{el}$, were $\hat\beta_\text{el}$ is the standard Bell operator defined by the inequality (\ref{elineq}) and the measurement operators by Alice and Bob:
\begin{eqnarray}
\label{elop}
\hat\beta_\text{el}&=&A_1\otimes(B_1+B_2-B_3-B_4)\nonumber\\
&+&A_2\otimes(B_1-B_2+B_3-B_4)\nonumber\\
&+&A_3\otimes(B_1-B_2-B_3+B_4) .
\end{eqnarray}
Finding the largest quantum violation of the elegant Bell inequality~\eqref{elineq} corresponds to maximising (minimising) the largest (smallest) eigenvalue of $\hat\beta_\text{el}$ ($\bar{\beta}_\text{el}$) over all quantum observables $A_i$ and $B_j$ by Alice and Bob.

Using the fact that $(A_i)^\dagger A_i=(B_j)^\dagger B_j =\identity$, it is easy to see that $\bar{\beta}_\text{el}$ admits a sum of squares (SOS) decomposition in the form $\bar{\beta}_\text{el}=(\sqrt 3/2)\sum_{\lambda}P^{\dagger}_{\lambda}P_{\lambda}$, where $P_{\lambda}$ are linear combinations of the operators $A_i$ and $B_j$ as follows
\begin{align}
P_1 &= (A_1+A_2+A_3)/\sqrt 3 - B_1\nonumber\\
P_2 &= (A_1-A_2-A_3)/\sqrt 3 - B_2\nonumber\\
P_3 &= (-A_1+A_2-A_3)/\sqrt 3 - B_3\nonumber\\
P_4 &= (-A_1-A_2+A_3)/\sqrt 3 - B_4,
\label{sos}
\end{align}

This implies that $\bar{\beta}_\text{el}$ is positive semidefinite, hence $\beta_\text{el} = 4\sqrt 3\identity - \bar{\beta}_\text{el}\le 4\sqrt 3\identity$. Therefore, the maximum quantum value of $\beta_\text{el}$ is upper bounded by $4\sqrt 3$. Since this value is attained with a quantum realization presented above, we have a new Tsirelson-type bound for quantum correlations. Note that by applying the methods of Refs.~\cite{ekb,julio} one gets the same upper bound of $4\sqrt 3$ on the maximum quantum value of the Bell inequality $\beta_\text{el}$.

\section{Certifying global randomness}
Here we give a detailed description of the Bell setup used to generate global randomness exceeding 2 bits from two-qubit entangled states. Note that according to the main text it is not possible to certify more than 4 bits of global randomness from any two-qubit state. Since the observed statistics provide an upper bound to the intrinsic randomness of the setup, in order to obtain 4 bits, we have to find a pair of 4-outcome extremal measurements on a two-qubit state which yield $P(a,b)=1/16$ for all results. To this end, let us pick the maximally entangled state $|\phi^+\rangle=(|00\rangle+|11\rangle)/\sqrt{2}$ and Alice's and Bob's rank-1 extremal POVM's as follows:
\begin{align}
\label{RaRb}
R_a =& (1/4)(\identity + \vec u_a\cdot\vec\sigma)\nonumber\\
R_b =& (1/4)(\identity + \vec v_b\cdot\vec\sigma),
\end{align}
where the four outputs are $a,b=\{1,2,3,4\}$, and $\vec\sigma$ denotes the vector of three Pauli matrices $(\sigma_x,\sigma_y,\sigma_z)$.

Alice's Bloch vectors $\vec u_a$ are distributed as
\begin{equation}
\begin{tabular}{llrrr}
$\vec u_1=$ & $c ($ & $-1,$ & $\delta,$ & $0$)\\
$\vec u_2=$ & $c ($ & $-1,$ & $-\delta,$ & $0$)\\
$\vec u_3=$ & $c ($ & $1,$ & $0,$ & $-\delta$)\\
$\vec u_4=$ & $c ($ & $1,$ & $0,$ & $\delta$),\\
\end{tabular}
\label{ua}
\end{equation}
while Bob's Bloch vectors $\vec v_b$ look as follows
\begin{equation}
\begin{tabular}{llrrr}
$\vec v_1=$ & $c ($ & $-\delta,$ & $0,$ & $-1$)\\
$\vec v_2=$ & $c ($ & $\delta,$ & $0,$ & $-1$)\\
$\vec v_3=$ & $c ($ & $0,$ & $-\delta,$ & $1$)\\
$\vec v_4=$ & $c ($ & $0,$ & $\delta,$ & $1$)\\
\end{tabular}
\label{vb}
\end{equation}
where $c=1/\sqrt{1+\delta^2}$ is a normalization factor and $\delta$ is a small positive constant. Note that the measurements defined above fulfill two necessary conditions to approximate the value of $P(a,b)=1/16$ for all results. (i) Vectors $\vec u_a$ and $\vec v_b$ become orthogonal when $\delta$ tends to zero. (ii) Both sets of vectors $\vec u_a$ and $\vec v_b$ span the three-dimensional space (i.e. the POVM elements are complex valued).

Let us now construct a 7-setting 2-outcome Bell inequality which reads
\begin{align}
\label{forkineq}
&\beta_{\text{fork}}=E_{11}+E_{22}+E_{33}\nonumber\\
& + E_{14}+\delta E_{24}+ E_{15}-\delta E_{25}\nonumber\\
& -E_{16}+\delta E_{36} -E_{17}-\delta E_{37} \nonumber\\
 &+ \delta E_{41}+E_{43} -\delta E_{51}+E_{53}\nonumber\\
 & + \delta E_{62}-E_{63}
                  + -\delta E_{72}-E_{73}\le 11,
\end{align}
where $0<\delta<1$. It is known that for a two-setting Bell inequality $E_{11}+\delta E_{21} + E_{12}-\delta E_{22}$ the quantum maximum is $2\sqrt{1+\delta^2}$. Using this result, it is straightforward to establish the upper bound of $3 + 8\sqrt{1+\delta^2}$ on the maximum quantum value of the above inequality $\beta_{\text{fork}}$. This value can in fact be attained by using the state $|\phi^+\rangle$ and observables $A_1=B_1=\sigma_x$, $A_2=-B_2=\sigma_y$, $A_3=B_3=\sigma_z$, and
$A_{x+3}=-\vec v_x\cdot\vec\sigma$ for $x=1,\ldots,4$ and $B_{y+3}=-\vec u_y\cdot\vec\sigma$ for $y=1,\ldots,4$, where the Bloch vectors above are defined according to equations~(\ref{ua},\ref{vb}). In this construction, the diagonal term $E_{11}+E_{22}+E_{33}$ forces Alice and Bob to choose correlated measurements, whereas the other parts of the inequality are used to steer Alice and Bob's measurements to the ones defined by the Bloch vectors~(\ref{ua},\ref{vb}).

We now introduce a four-outcome measurement on Alice's and Bob's side, which will be used to generate the global randomness. With the help of the above inequality~(\ref{forkineq}), we define a modified Bell inequality similarly to equation~(\ref{model}) in the main text as follows
\begin{align}
\label{modfork}
\beta'_{\text{fork}} &= \beta_{\text{fork}} - k\sum_{i=1}^4{P(a=i,b=+1|x=8,y=i+3)}\nonumber\\
&-k\sum_{i=1}^4{P(a=+1,b=i|x=i+3,y=8)}\le 11,
\end{align}
where $k$ is a strictly positive constant. The subtracted terms above will take care of aligning Alice and Bob's four-outcome measurements~(\ref{RaRb}) in the correct way (i.e. force them to take a form having Bloch vectors (\ref{ua}) and (\ref{vb})). Hence, in principle the probability $P(a,b)=1/16$ can be approximated for all $a,b$ pairs provided $\delta$ tends to zero.

Solving the above problem at order $1+AB$ of the NPA hierarchy we get a 121-dimensional moment matrix. By taking $\delta=0.0676946$ and $k=1$, the problem in this level can be solved on a desktop, and our technique certifies $2.8997$ bits of global randomness.


\vfill

\begin{thebibliography}{}
\bibitem{bell} J. Bell, Physics, \textbf{1}, 195 (1964).
\bibitem{rmpbell} N. Brunner, D. Cavalcanti, S. Pironio, V. Scarani and S. Wehner, Rev. Mod. Phys. \textbf{86}, 419 (2014).
\bibitem{nature} S. Pironio \emph{et al.}, Nature {\bf 464}, 1021 (2010).
\bibitem{njp} O. Nieto-Silleras, S. Pironio and J. Silman, New J. Phys. {\bf 16}, 013035 (2014).
\bibitem{scaranibancal} J.-D. Bancal, L. Sheridan and V. Scarani, New J. Phys. \textbf{16}, 033011 (2014).
\bibitem{colbeck} R. Colbeck, PhD Thesis University of Cambridge, arXiv:0911.3814 (2006).
\bibitem{vv} U. Vazirani and T. Vidick,  arXiv:1111.6054.
\bibitem{millershi} C. A. Miller and Y. Shi,  arXiv:1411.6608.
\bibitem{npa} M. Navascu\'es, S. Pironio and A. Ac\'\i n, Phys. Rev. Lett. {\bf 98}, 010401 (2007); New J. Phys. {\bf 10}, 073013 (2008).
\bibitem{amp} A. Ac\'\i n, S. Massar and S. Pironio, Phys. Rev. Lett. \textbf{108}, 100402 (2012).
\bibitem{maxrand} G. de la Torre \emph{et al.}, Phys. Rev. Lett. \textbf{114}, 160502 (2015).
\bibitem{acin} C. Dhara, G. Prettico and A. Ac\'\i n, Phys. Rev. A \textbf{88}, 052116 (2013).
\bibitem{scarani} V. Scarani, unpublished, see~\cite{acin}.
\bibitem{gisin} N. Gisin, arXiv:quant-ph/0702021v2 (2007).
\bibitem{tamas} T. V\'ertesi and E. Bene, Phys. Rev. A {\bf 82}, 062115 (2010).
\bibitem{extrpovm} G.~M. D'Ariano, P. Lo Presti and P. Perinotti, J. Phys. A: Math. Gen. \textbf{38} 5979, (2005).
\bibitem{chsh} J. F. Clauser \emph{et al.}, Phys. Rev. Lett. \textbf{23}, 880 (1969).
\bibitem{dhara} C. Dhara, G. Prettico and A. Ac\'{i}n, Phys. Rev. A \textbf{88}, 052116 (2013).
\bibitem{stchsh} M. McKague,  arXiv:1006.2352.
\bibitem{mosca-mckague} M. McKague and M. Mosca, arXiv:1006.0150 (2010).
\bibitem{ekb} M. Epping, H. Kampermann and D. Bru{\ss}, Phys. Rev. Lett. {\bf 111}, 240404 (2013).
\bibitem{julio} J. I. de Vicente, arXiv:1502.01876 (2015).
\bibitem{ncpol2sdpa} P. Wittek, arXiv:1308.6029 (2013).
\bibitem{sdpagmp} M. Nakata, In Proceedings of IEEE International Symposium on Computer-Aided Control System Design (2010).

\bibitem{gallego} R. Gallego et al, New J. Phys. 16, 033037 (2014).

\end{thebibliography}
\end{document}